# Financial Option Insurance

**Qi-Wen WANG**[a] and **Jian-Jun SHU**[b*]

[a]School of Business, Shanghai DianJi University, 1350 Ganlan Road, Lingang New City, Pudong New District, Shanghai 201306, People's Republic of China
[b]School of Mechanical & Aerospace Engineering, Nanyang Technological University, 50 Nanyang Avenue, Singapore 639798

## ABSTRACT

The option is a financial derivative, which is regularly employed in reducing the risk of its underlying securities. However, investing in option is still risky. Such risk becomes much severer for speculators who utilize option as a means of leverage to increase their potential returns. In order to mitigate risk on their positions, the rudimentary concept of financial option insurance is introduced into practice. Two starkly-dissimilar concepts of insurance and financial option are integrated into the formation of financial option insurance. The proposed financial product insures investors' option premiums when "misfortune" befalls on them. As a trade-off, they are likely to sacrifice a limited portion of their potential profits. The "loopholes" of prevailing financial market are addressed and the void is filled by introducing a stable three-entity framework. Moreover, a specifically designed mathematical model is proposed. It consists of two portions: the business strategy of matching and a verification-and-modification process. The proposed model enables the option investors with calls and puts of different moneyness to be protected by the issued option insurance. Meanwhile, it minimizes the exposure of option insurer's position to any potential losses.

## 1. Introduction

Contemporary financial market is always associated with the increasing volatility and uncertainty. The market participants, no matter individual investors or institutional traders, are unavoidably exposed to the *risk*[1] induced by the random events, which is generated by economic environment (central bank monetary policies, inflation, business cycles, global financial events, critical economic data announcement, *etc.*). Therefore, there is always a saying among traders: "Trading may have princes, but nobody stays a king" [1]. In order to beat the market, agile and accurate anticipations based upon real economic activities are the essential skills for individual investors, fund managers, and corporate financial managers. However, by contrasting the past predictions with respect to the actual financial market movements, the result remains pessimistic [2].

The financial market hates uncertainties, as any inappropriate strategies can simply spell catastrophic consequences to their retirement account, representing clients as well as organizations because of a considerable amount of money involved. Nowadays, due to the highly globalized nature of all large economies, any ominousness in one sector would be quickly disseminated, and ultimately become a global nightmare.

* Correspondence should be addressed to Jian-Jun SHU, mjjshu@ntu.edu.sg

[1] ***Risk***: In financial market, five types of *risks* are generally encountered by market participants, including market *risk*, opportunity *risk*, inflationary *risk*, credit *risk* and liquidity *risk*.

Therefore, in the perspective of market participants, it is always ideal to seek out an avenue to reduce the potential *risk* of their *portfolio*. Even in the perspective of the government, it is always optimal to introduce one additional entity to contemporary financial market, which helps to redistribute and regulate the *risk* among entities.

With the introduction of equity *option* into the prevailing financial market, the variety of trading strategies is significantly enriched. Constructing a well-diversified *portfolio*[2], which balances out or limits the exposure of underlying asset to any potential market fluctuations [3], becomes a plausible solution.

Nowadays, the most actively traded financial *derivative*[3] is *option*, which represents a contract sold by an *option* seller (*writer*) to an *option* buyer (*holder*) in exchange for the credit (*an option premium*[4]). Such a contract offers the *holder* a right, but not an obligation, to either buy (*call*) or sell (*put*) a specific financial instrument (*underlying security*) at a specific price (*strike price*[5]/*exercise price*) on (or prior to) a predefined date (*maturity date*). If the *option* is *in-the-money* (*ITM*[6]) on (or prior to) its *maturity date*, the *option holder* chooses to exercise the right (*option*), therefore, the *option writer* is obliged to deliver the underlying asset to the *option holder* (*call option*) or accept the delivery of the underlying asset from *option holder* (*put option*) at *strike price*. However, if the executional condition is unfavorable to the *option holder* prior to (or on) the *maturity date*, that is, the *option* is either *at-the-money* (*ATM*[7]) or *out-of-the-money* (*OTM*[8]), he has the right to give up the right and lets it be expired worthlessly on the *maturity date*, which directly incurs a loss of the entire *option premium*.

In reality, constructing an appropriate *portfolio* by employing *option* as an effective hedge against potential *risk* demands the investors with specialized skills like *risk* identification and quantifying. Unfortunately, the majority of investors and even financial professionals are not the true beneficiary of such a sophisticated method. Any misestimation may significantly cut down their returns or even exacerbate the losses due to the cost of hedging.

Moreover, in addition to the motivation of utilizing *option* as a hedge to limit the potential losses of a position against adverse market movements, *option* is frequently adopted as the means of leverage to improve investors' potential gains by speculating on the market movements of an *underlying security*. By adopting such a strategy, the *speculators* are exposing their position completely unprotected to the market *risk*. As is known to all, the monetary value of the financial *option* consists of two parts, known as intrinsic value and time value. The intrinsic value is directly related to the spot price and volatility of its underlying asset, while its time value vanishes once it approaches the specified expiry date [3]. Therefore, from a certain perspective, speculating in *options* can be deemed as one of the most aggressive trading approaches. It is because it demands the *speculators* to anticipate the price movements of the underlying asset accurately within a specified time horizon, that is, prior to the *maturity date*. The *risk* of *options* has been technically reported in [4–14].

According to consolation hypothesis [15], it is reasonable to perceive insurance compensation as a token of consolation in monetary decisions. Such willingness is significantly magnified with the increased level of market uncertainties. Built upon this hypothesis, an accessible trading scenario for *option hedgers* and/or *speculators*, who agree to sacrifice the limited portion of their potential profits in exchange for protection, is suggested in this paper. The developed financial instrument is named as "financial *option* insurance".

---

[2] ***Portfolio***: A collection of negatively correlated financial instruments.
[3] ***Derivative***: A financial instrument whose value is merely depending on its underlying asset.
[4] ***Option premium***: The market price of *option* during the transaction between *option writer* and *holder*.
[5] ***Strike price***: The price at which the *option holder* buy or sell *underlying security* as specified in the contract.
[6] ***In-the-money*** (***ITM***): The *strike price* of *call option* is lower than the spot price of its *underlying security* or the *strike price* of *put option* is higher than the spot price of its *underlying security*.
[7] ***At-the-money*** (***ATM***): The *strike price* of *call/put option* is close or equal to the spot price of its *underlying security*.
[8] ***Out-of-the-money*** (***OTM***): The *strike price* of *call option* is higher than the spot price of its *underlying security* or the *strike price* of *put option* is lower than the spot price of its *underlying security*.

Unlike the conventional way of hedging, which is achieved by combining negatively correlated financial *derivatives* in the formation of financial *portfolio*, financial *option* insurance can be issued by a completely independent institution. It works very alike to an insurance company dedicated to the *option* market. The proposed financial instrument integrates two starkly-dissimilar concepts of insurance [16] and financial *options* [17]. This allows any *option holders* to claim partial compensation to mitigate their cost of the *option* if they have purchased the insurance contract by paying an additional *insurance premium* during the transaction of the corresponding *option*. Financial *option* insurance can be sold by an existing financial institution as an extra service to either interested investors or a dedicated company. For the sake of convenience, in the remaining part of this paper, such an organization, which provides the service of financial *option* insurance, is simply referred to as the *third entity* or *option insurer*.

In order to guarantee that the business holds a position, which generates the positive cash flows for each transaction, and is completely independent of the financial market trend, it is always ideal for option insurer to seek a pair of *option* investors with totally reversed market expectations (*i.e.*, a *call* and a *put* of the identical *strike price* and *expiration date*). However, on most occasions, it is not realistic to have the equal-sized group of the *call option* and *put option* investors entering into the *matching system* of the *third entity*. It is because the price of an exchange traded financial instrument fluctuates at any moments, which reflects different expectations about the future trend of the corresponding asset as well as various applied strategies of market participants. In order to enable the investors with *calls* and *puts* of different moneyness to be protected by the issued *option* insurance, a specifically designed mathematical model is proposed in this paper. It has two portions: the business strategy of matching and a verification-and-modification process. The proposed model enables the *option* investors with *calls* and *puts* of different moneyness to be allocated as paired investors. Meanwhile, it minimizes the exposure of *option* insurer's position to any potential losses. The business strategy of matching is analogous to the generalized Tian Ji's horse racing strategy [18–20] and the Nobel prize-winning stable allocation theory [21,22]. In the end, the novelty of financial *option* insurance is elaborated in three aspects including market acceptance, *risk* profile and profitability, and positive market effect.

The fundamental business model of financial *option* insurance company is introduced in Section 2, which includes the detailed insurance policies and working principles of the *option insurer*. Three deliberately selected examples are employed to demonstrate the feasibility of introducing the financial *option* insurance into the prevailing market. To enhance the functionality of financial *option* insurance, a specifically designed mathematical model, which comprises the business strategy of matching and a verification-and-modification process, is proposed in Section 3. In Section 4, the potentiality of financial *option* insurance is discussed in terms of three aspects, including market acceptance, *option insurer risk* profile and profitability, and simulated positive market effect with the involvement of *option insurer*.

# 2. Business Model of Financial Option Insurance

The fundamental business model for a financial *option* insurance company, namely the *third entity*, is described in terms of insurance policies and the working principles. Moreover, three examples of fabricated trading scenarios are employed to demonstrate the feasibility of introducing financial *option* insurance into prevailing financial *option* market.

## I. Insurance policies and working principles

The purpose of insurance is to allow a *third entity* to sell financial *option* insurance to *option* buyers, mainly for those who longed a naked *option* position, in protecting their *options* against adverse market movements. The insured investors are entitled to claim reimbursement from the *third entity* when the corresponding bet goes wrong.

The insurance policies can be briefly summarized in terms of seven (7) major statements as indicated below:

1. The insurance contract is available for both parties, *call* and *put option* buyers, at any time during the transaction of the *option* before its *maturity date* (or *expiration date*).
2. Insurances are sold in pairs, *i.e.*, when there is a matching in the *strike price* and *expiration date* of the corresponding *call* and *put option* buyers. The matching mechanism of different *strike prices* is elaborated in Section 3.
3. Total payable *insurance premiums* are divided and shared by paired *option* investors at a predetermined percentage (known as a yardstick). The yardstick is determined by a specifically designed pricing structure.
4. The *third entity* only compensates/reimburses the party, whose *option* is *OTM* at maturity. The amount of reimbursement was stipulated in the insurance contract during the time it was issued by the *third entity*.
5. *Option* insurance contracts are standardized as its underlying *option* is a *vanilla option* (*i.e.*, a standardized *option* which can be traded on an exchange). Moreover, *option* insurance and its underlying *option* are not bounded together. Therefore, the insurance contract and the underlying *option* can be separately traded on the *secondary market of insurance contracts* (*i.e.*, a platform hosted and regulated by the *third entity*) and on an *option* exchange, respectively, within the life time of the contract. A commission fee is charged by the *option insurer* upon each successful transaction of ownership of an insurance contract. The role of *option insurer* is limited on maintaining the day-to-day operations and providing related services to the potential customers enrolled in this platform.
6. As long as the ultimate insurance contract *holder* can provide the evidence, which indicates holding the same amount of unexercised *option* of the identical type (*call* or *put*), *strike price*, and *expiration date*, as stipulated in the *option* insurance contract, which is *OTM*, to maturity, without subjected to the compensation of the *third entity*, the *third entity* reimburses the ultimate *holder* the same amount as stipulated in the insurance contract.
7. If the *option* is *ITM* or *ATM* at maturity, due to any other reasons, the *option holder* was unable to exercise the *option*. Or the insurance contract *holder* cannot provide the evidence of holding the specified *option* to maturity. The *third entity* does not reimburse the insurance contract *holder*, profiting the entire pre-collected *insurance premium*.

Based on the established insurance policies, it is able to represent the entire business model in terms of a schematic representation as illustrated in Figure 1. (1) Long *call* and long *put* investors purchased the corresponding *options* from the *option* exchange. (2) If they have willingness to sacrifice the limited portion of their potential profits in seeking de facto insurance protection, they can either entering into the *matching system* by indicating their existing position (*calls* or *puts*) to the *third entity* and being allocated by *matching system* to open a new insurance contract or directly purchasing an insurance contract of the identical *underlying security*, *option* type, *strike price*, and expiry date (without indicating their *option* positions to the *option insurer*) from the *secondary market of insurance contracts* hosted by the *third entity*. (3) Once the *matching system* completes the allocation process, the *insurance contract issuing unit* determines the *insurance premium* for paired investors based on a specific yardstick and drafts the corresponding reimbursement clauses. The completed proposal is sent back to investors for acceptance. (4) If both paired investors accepted the clauses as specified in the proposal, the *third entity* charges them *insurance premium* stipulated in the proposal, and the insurance contract (which specifies the *option*, *option* type, *strike price*, expiry date, number of *option* covered, and reimbursement clauses) is issued to these investors. (4a) If any party of paired investors refused to accept the conditions as specified in the proposal, they are sent back to the *matching system* and waiting for another round of allocation, or alternatively, they may seek protections from the *secondary market of insurance contracts*. (5) The insured *option* investors are allowed to hold their insurance contracts until the *maturity date*. Alternatively, (5a) they may trade their insurance contracts in the *secondary market of insurance contracts*, a trading platform hosted and regulated by the *third entity*. The role of *option insurer* is limited on maintaining the day-to-day operations and providing related services associated with this platform. It does not engage in the trading (buying or selling) of the insurance contracts directly, and hence, insulates its potential *risk* caused by the unpredictable price movements of the insurance contracts. As a reward, the *third entity* charges a commission fee upon each transaction of the ownership of the insurance contract between the insurance contract buyer and seller in the secondary market (a certain fixed percentage of the total transaction value of the insurance contract). The spot price of the insurance contract is mainly determined by the supply and demand of the corresponding insurance contract, which is influenced by the instant price of the underlying *option*. It is important to note that traders (with or without an existing *option* position) are permitted to participate in this secondary market. (5b) The insured *option* investors sell their insurance contracts in the secondary market prior to expiry is either because they sold their *option* or they are confident about that their *option* can remain *ITM* (*i.e.*, failure to fulfill the reimbursement condition) prior to the *maturity date*. On such occasion, the *secondary market of insurance contracts* provides a platform for them to mitigate their hedging cost or even improve their potential returns. The insurance contracts can be repeatedly traded on the same platform among investors within its lifetime. (6) On the *maturity date* of the insurance contract, the *reimbursement unit* of the *third entity* notifies the ultimate insurance contract *holder*, whose *option*, as specified in the insurance contract, is *OTM*. As long as the insurance contract *holder* is able to submit the evidence of holding the same amount of expired *option* as stipulated in the insurance contract, the *third entity* reimburses the *holder* the amount, as specified in the insurance contract, after verifying the information. After reimbursing the corresponding option investor, the relevant information is recorded and the insurance contract is terminated by the *third entity*. (6a) The corresponding insurance contract is subjected to the termination for any invalid requests for reimbursement, that is, (i) the insurance contract specifies that the underlying *option* is *ITM* or *ATM*; (ii) the insurance contract *holder* fails to provide evidence, which indicates holding the same amount of specified *OTM option* to maturity; (iii) the unexercised *OTM option* is reimbursed by the *third entity*.

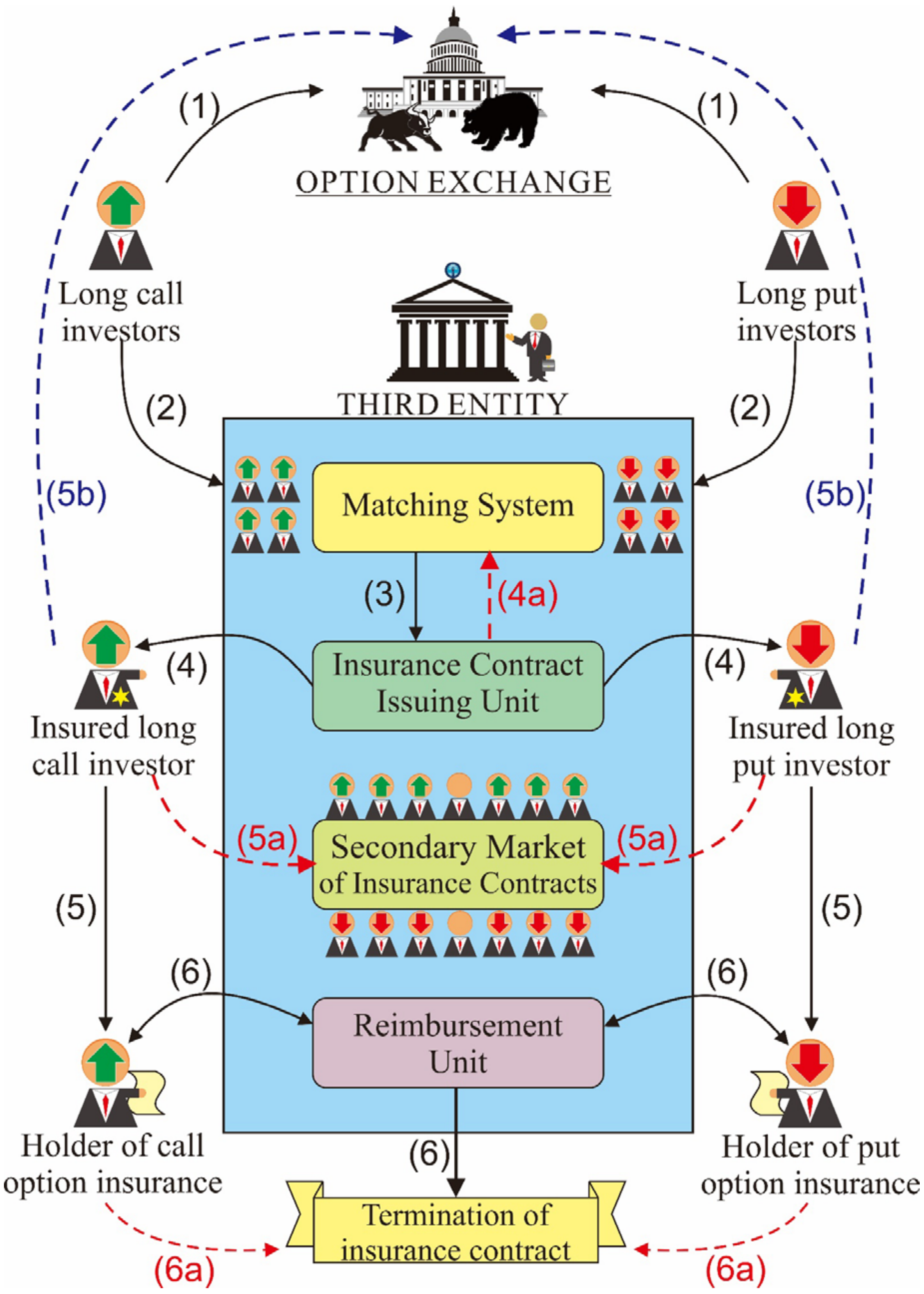

**Figure 1:** *A schematic representation of business on financial option insurance*

## II. Example 1: Scenario of holding insured option to maturity

In order to demonstrate the concept of insurance policies and the business model of financial *option* insurance, we deliberately select three possible trading scenarios. It begins with a considerably simple scenario: assuming that there are two *option* investors, let us say, investors **A** and **B**, both parties are interested in equity *IKEA*[9], and employ *option* as the means of leverage to improve their potential return on investment. However, these two investors have the completely reversed future expectations of *IKEA* – investor **A** expects the *IKEA* to go bullish, while investor **B** anticipates it to go bearish. Under the spot price of *IKEA*, which is $S_0$ = \$505, investor **A** decides to purchase 100 shares of *IKEA Feb 2013 500 Call @* premium price of C = \$24[10] and investor **B** decides to purchase 100 shares of *IKEA Feb 2013 500 Put @* premium price of P = \$15, assuming that both investors hold their position to maturity. Without purchasing the financial *option* insurance, investors **A** and **B** have potential to lose their entire *option premium* if the trend of market is unfavorable to either of them, which is described in Figure 2.

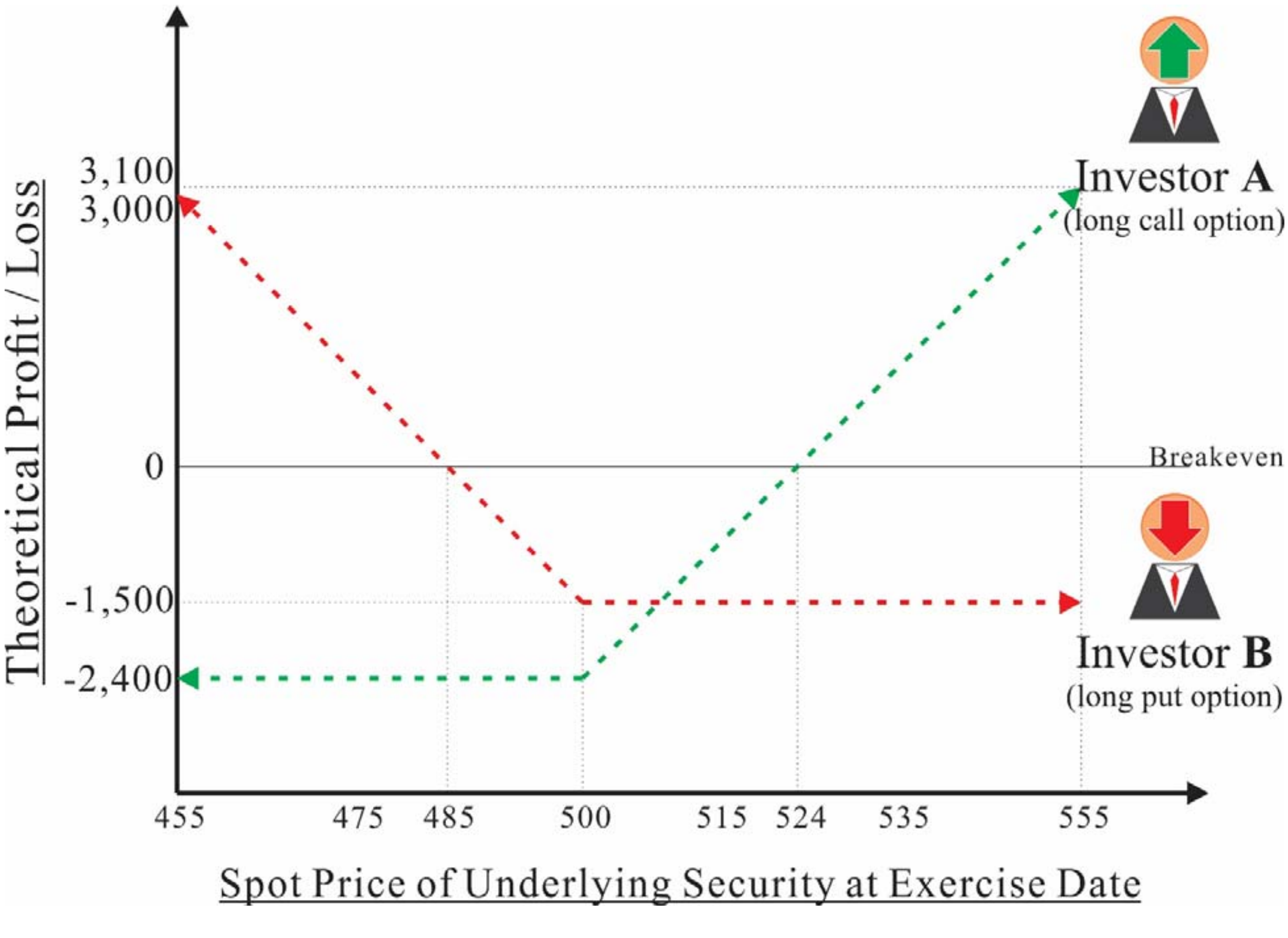

**Figure 2:** *Risk and reward graph for call and put option investors*

The situation can be significantly improved if both investors decide to purchase financial *option* insurance from the *third entity*, which is adopted to protect their originally naked positions. After opening the position (long *calls* or *puts*), they indicate their willingness to initiate a new insurance contract and the account information to the *third entity*. After examination process (a measure used to prevent any *option* investors from opening multiple copies of insurance contracts based up on the same *option*), they are allocated by the *matching system* of the *third entity* to be paired investors as their *options* have completed reversed expectation and the same amount of *calls* and *puts* with the identical *expiration date* and moneyness. The allocation is determined by a specifically designed matching mechanism, which is discussed in Section 3. The *third entity* charges an *insurance premium* (for financial *option* insurance) to each one of paired investors in order to insure them against their potential losses. The critical component in financial *option* insurance is to calculate on how much the *third entity* should charge for the *insurance premiums* in order to be profitable, while retaining attractiveness to *option* investors. The *third entity* operates in a manner that it charges the *insurance premium* based on a specific pricing structure. At current stage, assuming that the *third entity* utilizes 50% as the yardstick (this yardstick is subjected to the actual market conditions by including the factors like the liquidity, volatility, and future market expectations. In this example, 50% simply specify the anticipated price of the *underlying security* having the identical chance of going bullish and bearish), therefore, the *insurance premium* is 50% of the maximum value between the market prices of the corresponding *calls* and *puts*.

[9] ***IKEA***: Trading quote
[10] ***IKEA Feb 2013 500 Call @ 24***: Each *option* quote consists of five components – *option* symbol (*IKEA*), *expiration date* (Feb 2013), *strike price* (\$500), *option* type (*call option*), and *option premium* (\$24).

In this case, the *option premium* of the *call* ($24) is more than that of the *put* ($15), *i.e.*, $C > P$. Based on the given pricing structure, the *third entity* charges an *insurance premium* of $12[11] (= 50% max [$24, $15]) for undertaking the *risk* of insured *option* investors, *call* and *put option* investors. This amount $12 is split evenly (due to the 50% yardstick) between these two investors such that each one of them pays only $6 as an *insurance premium*. The entire transaction process of three-entity structure is outlined in Figure 3.

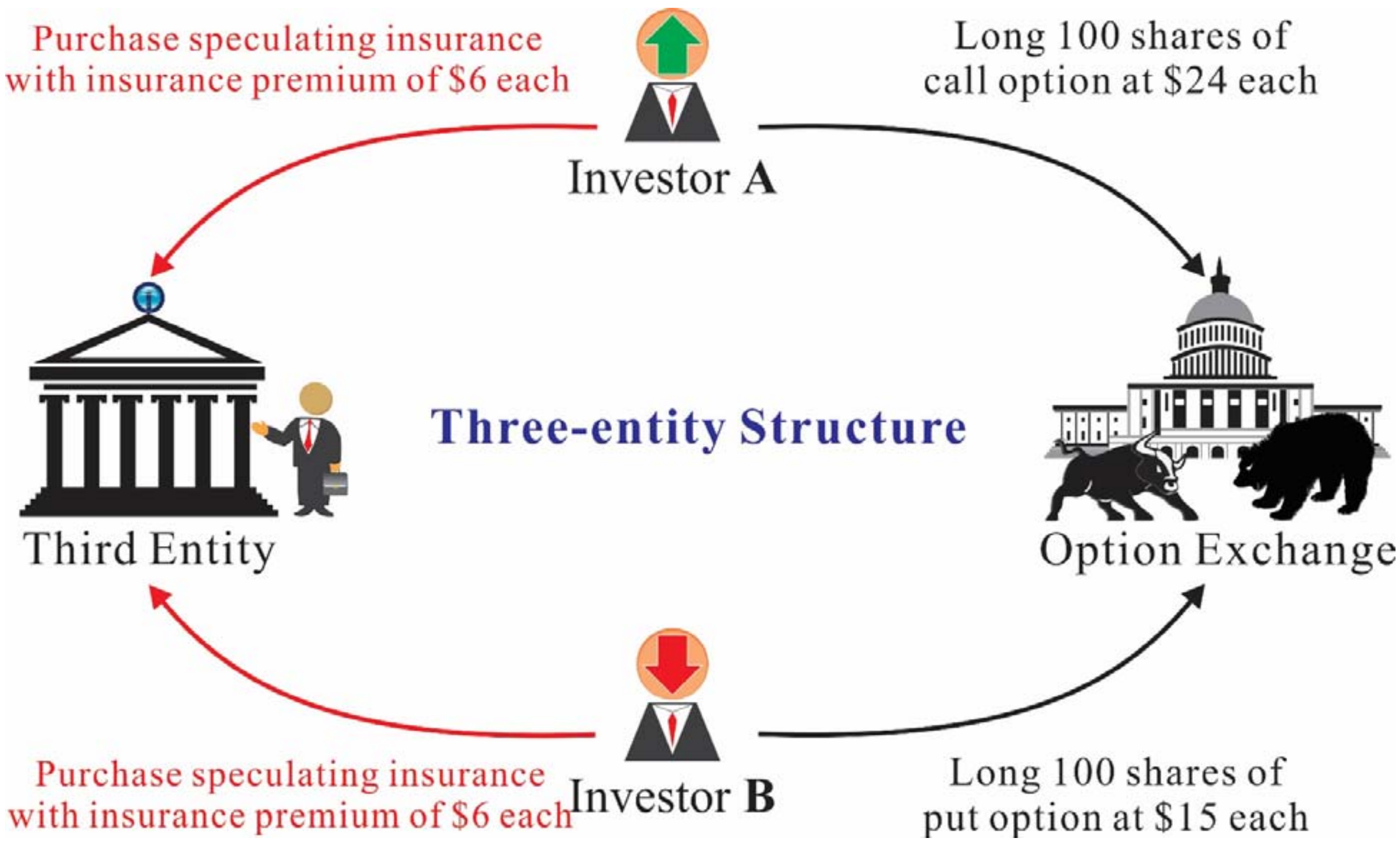

**Figure 3:** *Investors purchase exchanged-traded options and financial option insurance*

The main benefit of the financial *option* insurance, as showed in this trading scenario, is that the *risk* is spread among more entities, resulting in a lower *risk* for each participant.

Assuming that insured investor **A**'s prediction is right, the price of *underlying security IKEA* is increased to $555 (*i.e.*, $S_{t1}$ = $555) on the *maturity date* of the *option*. Under this scenario, insured investor **A**, who brought

[11] ***Insurance premium***: ½ × max (*option premium of call*, *option premium of put*)

a *call option* and expected the price of *underlying security* to rise, chooses to exercise his *call option* and gains a net profit of \$2,500 [= 100 × (\$555 − \$500 − \$24 − \$6)]. On the other hand, he is invalid to claim reimbursement from the *third entity* as "misfortunate" does not befall on him.

On the contrary, insured investor **B**'s prediction is wrong as he expected the declination of price of its *underlying security IKEA* when he longed the *put option*. Since his *put option* is *OTM* at maturity, it expires worthlessly by incurring a net loss of \$1500 to him (*i.e.*, the amount paid for longing the *put option*). Under this scenario, the "misfortunate" befalls on insured investor **B**. Therefore, he is entitled to receive a reimbursement of \$742.5 (= 100 × 50% \$15 × 99%) after deducting the service charge. The service charge was specified in the insurance contract when it was issued by the *third entity* (in this case, 1% of the entire reimbursement amount). The investor **B** only pays \$600 for his financial *option* insurance. As a consequence, he stands to gain a net profit of \$142.5 purely from the transaction between him and the *third entity*. In this situation, the *third entity* also makes a net profit of \$457.5.

Conversely, let us consider a completely reversed situation. The spot price of its *underlying security* plunges to \$455 (*i.e.*, $S_{t2}$ = \$455) at maturity. This time the "misfortunate" has befallen on insured investor **A**. His *call option* expires worthlessly. Therefore, he is valid to claim a reimbursement of \$1,188 (= 100 × 50% \$24 × 99%) from the *third entity*, whereas only \$600 is paid for the financial *option* insurance. As a consequence, he stands to profit \$588 purely from the transaction between him and the *option insurer*.

Under this scenario, insured investor **B** is not capable of claiming any reimbursement from the *third entity*. The *third entity* only earns a service charge of \$12 (*i.e.*, 1% of the entire reimbursement amount) and the remaining collected *insurance premium* is paid to the insured investor **A** as compensation.

In summary, both investors stand a chance to mitigate their cost of longing the corresponding *option* if they had purchased the *option* insurance from the *third entity*. Similarly, the *third entity* earns either \$12 or \$457.5, that is, an expected value (EV) of \$2.348 per share of financial *option* insurance if both scenarios have the equal chance of occurring. Hence, it can be clearly observed that the proposed financial *option* insurance creates a win-win situation among all entities involved, which is depicted in Figure 4.

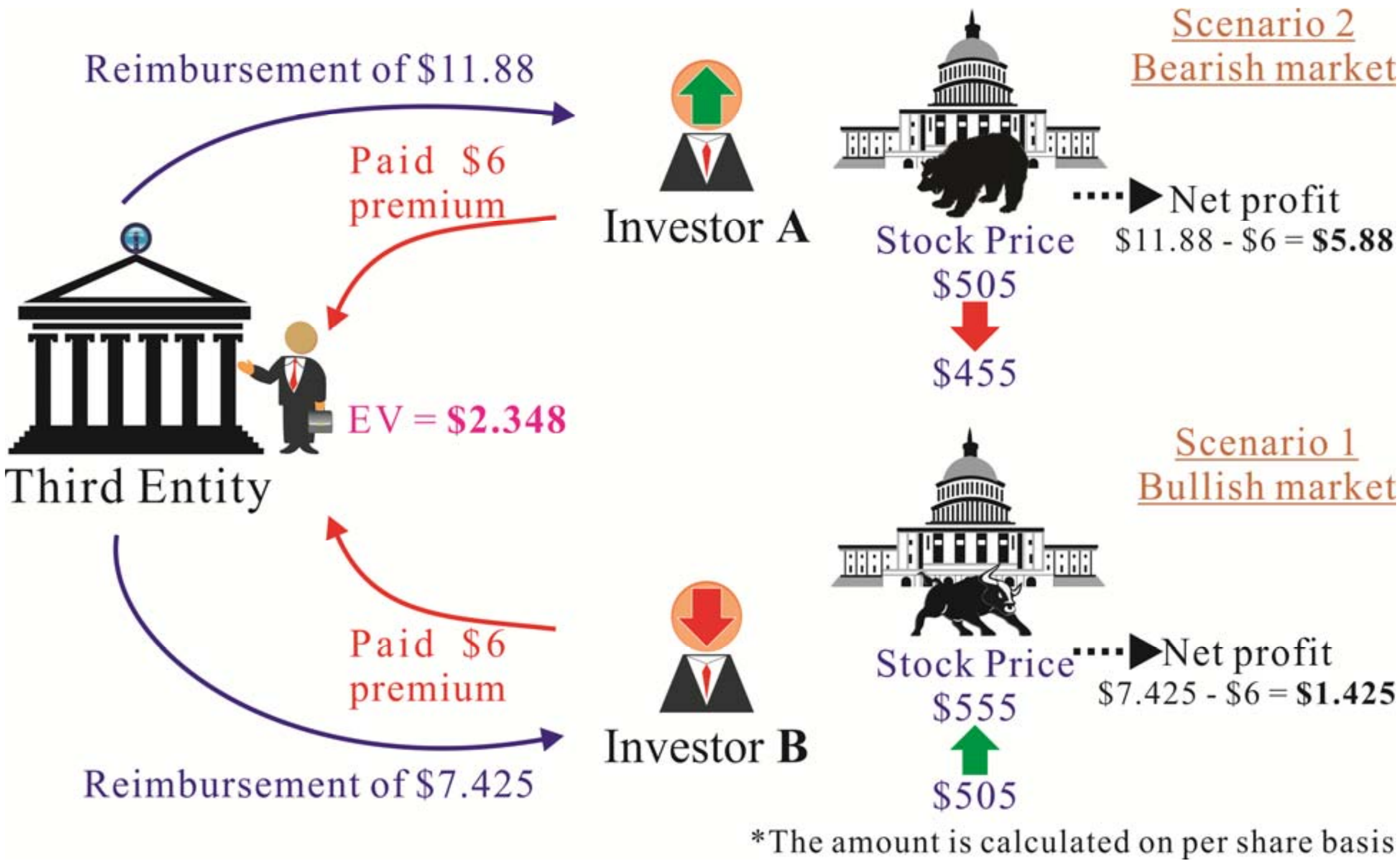

**Figure 4:** *A win-win situation for both investors and the third entity under two reversed scenarios*

Example 1 illustrates a relatively simple trading scenario, both investors, **A** and **B**, purchase an insurance contract to protect their position against unexpected market movements, and hold the position to its *maturity date*. It is utilized to emphasize the motivation of introducing the *third entity* into the contemporary two-entity framework, that is, it can be treated as an insurance company. Financial *option* insurance is applicable when the market price of the *option* is divulged as investor approaches the *third entity* to purchase the insurance. In addition, financial *option* insurance utilizes the essence of the insurance industry to establish a three-entity framework by creating a win-win situation for all entities involved, whereas *risk* can be quantified and reduced through spreading over more market participants. The *risk* and reward for investors, **A** and **B**, with and without the involvement of financial *option* insurance, are compared in Figure 5. Financial *option* insurance is acting as a new financial instrument to spread the initial *risk* among more entities, resulting in a lower *risk* for each market participant.

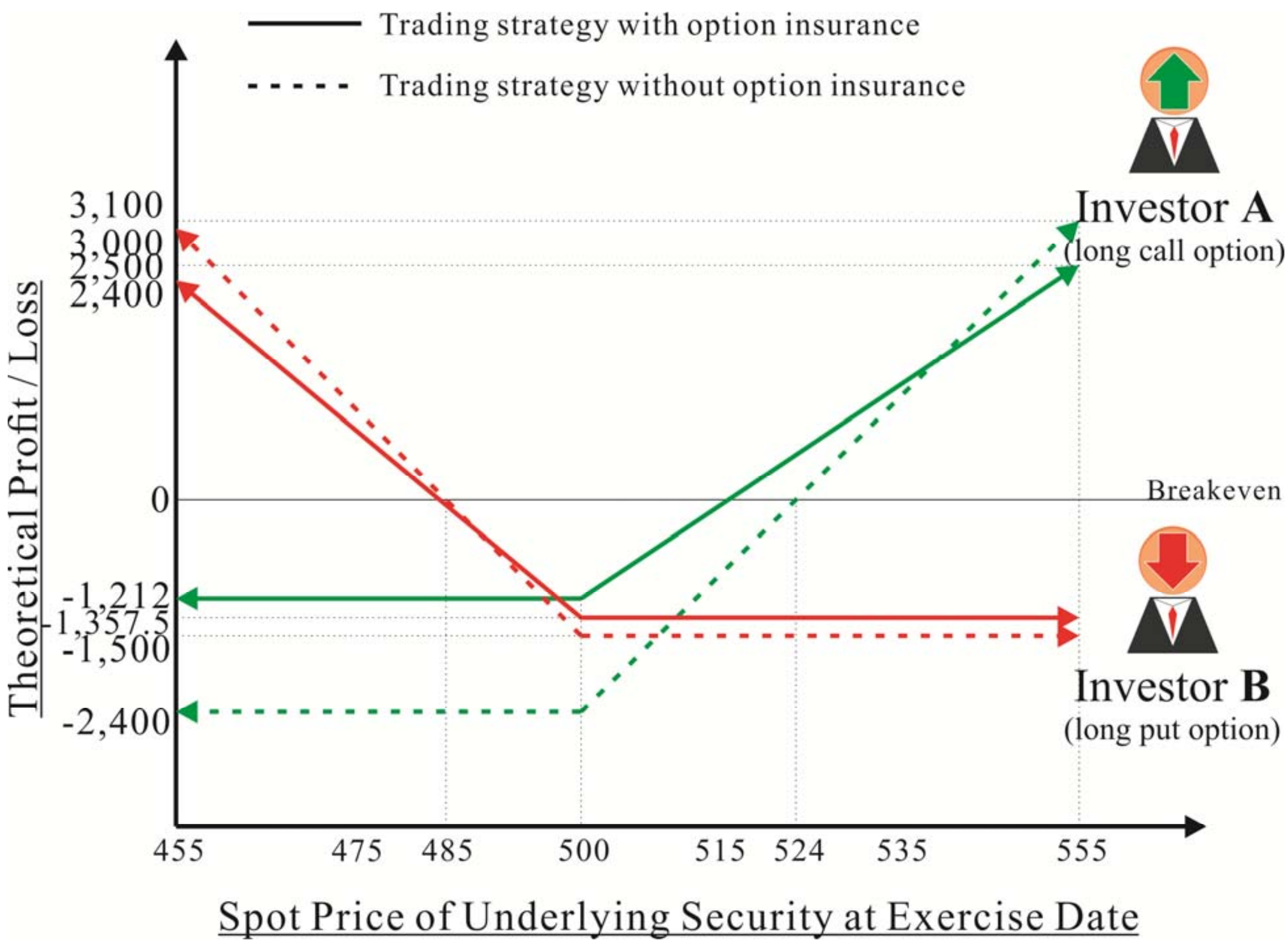

**Figure 5:** *Risk and reward graph for investors with and without financial option insurance*

## III. Example 2: Scenario of holding insured portfolio to maturity

Example 1 provides an elementary trading scenario, which consists of a long position on equity *option* protected by the corresponding insurance contract. It is proposed to spotlight the ability of the *third entity* in redistributing the *risk* among entities. However, the employment of financial *option* insurance is not only restricted in such a simple trading scheme. In reality, financial *option* insurance can be combined with an existing *portfolio*, as long as it contains a long position in *option*, to result in the formation of a new *portfolio* which is capable of handling a distinct situation. Example 2 demonstrates the feasibility of introducing the financial *option* insurance into an existing *portfolio*, which by itself, balances out or reduces the exposure of the underlying asset to any unfavorable market movements.

The existing *portfolio* is an implementation of hedging strategy, which is frequently adopted by financial professionals to reduce the exposure of their position to particular *risk*, mainly the market *risk*, which they may encounter, by taking positions through a negatively-correlated financial instrument that balances out or reduces the exposure of underlying asset to market fluctuations. Such a strategy generally involves one or several financial *derivatives*.

The hedging strategy generally involves two possible trading positions of one particular financial instrument, namely, long and short positions. Entering into a long position implies that the investor buys and owns an underlying asset. Entering into a short position implies that investor borrows one specific underlying asset from another entity (usually a brokerage firm) and sells it immediately in exchange for a credit in its trading account.

As the price of the underlying asset decreases, the investor can purchase it back at a lower price and return to the lender, profiting from the difference between selling and buying the price of the corresponding financial instrument. In order to open a short position, a *margin account*[12] is compulsory. The lender (broker) may at any time revise the value of the collateral securities (margin) to ensure the market value of the revised margin above that of collateral securities [17].

This trading scenario involves two conservative investors, let us say, investors **C** and **D**. Both investors are interested in the same stock *IKEA* with reversed future expectations. Investor **C** expects the *IKEA* to go bullish, while investor **D** anticipates it to go bearish. Therefore, the existing *portfolio* of investor **C** combines a long position on stock (*i.e.*, purchases 100 shares of *IKEA* @ spot price of $505) and a long position on *put option* (*i.e.*, purchases 100 shares of *IKEA Feb 2013 500 Put* @ premium price of $15). Investor **D**, on the other hand, integrates a short position on stock (*i.e.*, shorts 100 shares of *IKEA* @ spot price of $505) and a long position on *call option* (*i.e.*, buys 100 shares of *IKEA Feb 2013 500 Call* @ premium price of $24) in his *portfolio*. In both cases, by employing hedging strategy in their trading, the *risk* has been significantly limited. However, suppose both investors believe their anticipation is appropriate at the time they constructed their *portfolio*, but still desire to seek the protections to any conceivable market anomalies at a relatively lower cost, the introduction of financial *option* insurance into their constructed *portfolio* is feasible, as shown in Figure 6.

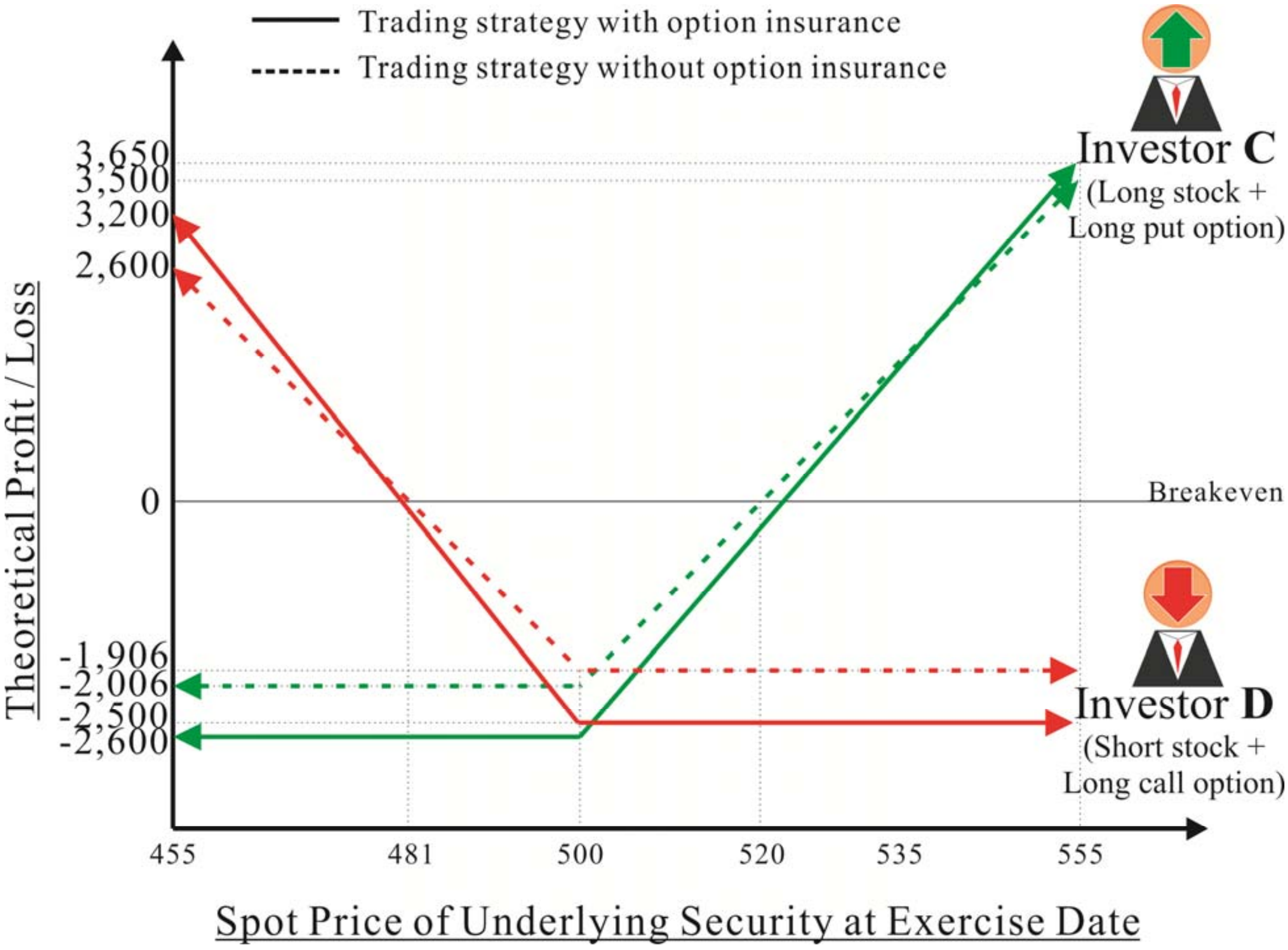

**Figure 6:** *Risk and reward graph for investors with and without financial option insurance*

[12] ***Margin account***: A *margin account* is a typical customer account. For each margin trade, the customer is allowed to put up a certain percentage of total cost of the trade amount in cash and the remaining in account as a form of deposit.

## IV. Example 3: Scenario of adjusting position in response to price movements

Examples 1 and 2 are relatively simple trading scenarios, assuming that both investors hold their *portfolio* until the *maturity date*. These two examples are employed to demonstrate the feasibility of utilizing financial *option* insurance to hedge the potential *risk* caused by market uncertainties. In reality, such "buy and hold" trading approach rarely occurs in real-life situations, especially with the involvement of an *option* in the selected *portfolio*. Investors may frequently adjust their position in response to the price movements of the *underlying security*. In this example, a more close to real-life trading approach is demonstrated by including various traders carrying different motivations.

The *portfolio* of investors **E** and **F** is quite identical to investors **A** and **B**, respectively. At the time, $t_0$, as specified in Figure 7, investor **E** expects the *IKEA* (the spot price of *IKEA* is $S_{t0}$ = $505 per share) to go bullish. He purchases 1,000 shares of *IKEA Feb 2013 500 Call* @ price of $C_{t0}$ = 24. At the same time, investor **F** expects *IKEA* to go bearish. He purchases 1,000 shares of *IKEA Feb 2013 500 Put* @ price of $C_{t0}$ = 15. At the same time, both investors apply for opening a new insurance contract and are successfully allocated by the *matching system* of the *third entity* to become paired investors, assuming the same yardstick of 50% is still employed by the *third entity*. Investor **E** pays $6,000 ($6 per share, a total of 1,000 shares of insurance contracts) as *insurance premium* in exchange for a reimbursement clauses of "reimbursing the insurance contract *holder* of unexercised *option IKEA FEB 2013 Call* a total amount of $12,000 ($12 per share) if the spot price of *IKEA* falls below $S_{tm}$ = $500". Investor **F** pays $6,000 as *insurance premium* in exchange for the reimbursement clause of "reimbursing the insurance contract *holder* of unexercised *option IKEA FEB 2013 Put* a total amount of $7,500 ($7.5 per share) if the spot price of *IKEA* rises above $S'_{tm}$ = $500".

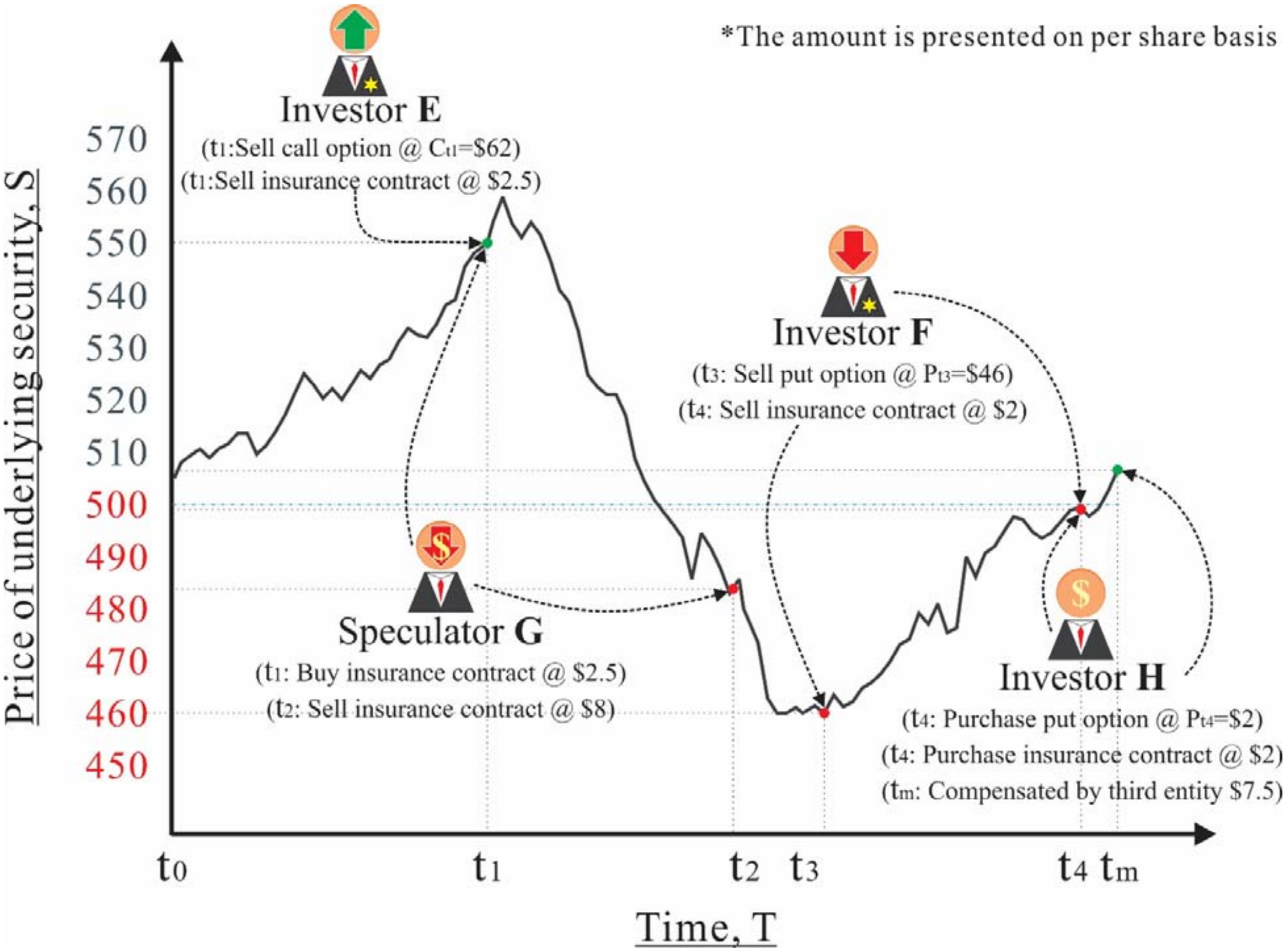

**Figure 7:** *Different trading strategies adopted by various traders along time*

At time, $t_1$, the instant stock price of *IKEA* soars to $S_{t1}$ = \$550 per share. Investor **E** chooses to lift the position by selling the *call option* at the *option* exchange at a price of \$62. In addition to that, investor **E** expects that the price of *IEKA* may experience some adjustments before the expiry date, but the chance of falling below \$500 per share (condition for the *third entity* to reimburse the insurance *holder*) is slim. It is worthwhile to sell all his insurance contracts in the *secondary market of insurance contracts* hosted by the *third entity* immediately at a price of \$2.5 (as long as there is still some time value of the insurance contract). The transaction cost is 1% of the market value of the insurance contract (assuming that the transaction cost charged by the *third entity* is 2%, which is split between the insurance contract buyer and seller). Therefore, total profit for investor **E** is \$34,475 [= (\$62 – \$24 – \$6 + \$2.5 – \$0.025) × 1,000 shares] before deducting the commissions charged by the broker.

At the same time, $t_1$, *speculator*[13] **G** believes that *IKEA* is overpriced as the market overreacts to its recently released annual report, which announces the earnings per share is increased by 8.5%. He anticipates that *IKEA* may quickly be subjected to an adjustment on its price, which drives the price to its average level of the time (known as *mean reversion*). In addition to that, there is a certain duration prior to the expiration of the insurance contract. It still stands a chance that the price of *IKEA* may fall below \$500. In short, he conceives that the contemporary price of \$2.5 per share is subject to mispricing. Therefore, he purchases 1,000 shares of insurance contracts from the secondary market at a price of \$2,525. With respect to the subsequent price movements of *IKEA*, his prediction is proved to be accurate. After hitting a high of \$558, the stock quickly plunges to a low of

[13] ***Speculator***: An investor engaged in risky financial transactions in attempts to maximize its potential profits from the underlying financial attributes in the market value of tradable goods.

\$486. At this time, the buyers reappear, and the price of *IKEA* jumps back to \$495. However, the downward momentum remains, and the price quickly descends to \$484. After the price of *IKEA* re-stabilizes at the time, $t_2$, *speculator* **G** decides to exit the position by selling the insurance contract at a price of \$8 per share. It is taking the consideration that his position is not properly hedged. And even if the price of *IKEA* falls to another low, the potential profit for him is restricted by the reimbursement clauses of *call option* insurance (theoretical maximal profit is \$12 per share). Total profit for *speculator* **G** is \$5,395 [= (\$8 – \$2.5 – \$0.08 – \$0.025) × 1,000 shares].

At time, $t_2$, investor **F** has the similar prediction to that of *speculator* **G**. In contrast to the position of *speculator* **G**, investor **F** holds a hedged position. Moreover, his potential gain is not limited to a certain fixed amount as that of *speculator* **G**. Therefore, he is holding the position until the price of *IKEA* drops to \$460. After the price reaches the new bottom at \$460, he decides to lift the position on a *put option* at the price of $P_{t3}$ = \$46, on time, $t_3$. However, he decides not to sell the *put option* insurance contract immediately as the market price is much lower than what he was paid on purchasing the insurance contract. He holds the *put option* insurance until time, $t_4$, which is very close to the due date of the insurance contract. In the end, he manages to sell the insurance contract in the secondary market at a price of \$2 per share. Total profit for investor **F** is \$26,980 [= (\$46 – \$15 – \$6 + \$2 – \$0.02) × 1,000 shares] before deducting the commissions charged by the broker.

At time, $t_4$, the instant price of *IKEA* rebounds back to \$499. The market price of *IKEA Feb 2013 500 Put option* is \$2. Meanwhile, the *put option* insurance is \$2 as well. Investor **H** realizes that it may be the opportunity to take very limited *risk* (the price of *IKEA* on expiry falling in the range between \$496 and \$500) in exchange for a possible profit of \$7.5 per share if the price of *IKEA* on maturity rises above \$500 ($S'_{tm}$ > \$500). Therefore, he purchases 1,000 shares of *IKEA Feb 2013 500 Put @* $P_{t4}$ = \$2, and 1,000 shares of the *put option* insurance. At the *maturity date*, the price of *IKEA* is $S_{tm}$ = \$506. Therefore, he is legal to claim reimbursement from the *third entity* as his insured *put option* expires worthlessly. Total profit for investor **H** is \$3,405 [= (\$7.5 × 99% – \$2 – \$2 – \$0.02) × 1,000 shares] before deducting the commissions charged by the broker.

By observing the profit of traders **E**, **F**, **G**, and **H**, it is possible to reach into a false conclusion that every party may be benefited by introducing the financial *option* insurance into their trading approach. This is not what we desire to advocate here. It is because all of these fabricated traders profit from their positions merely by choosing appropriate entering and exiting strategies, and fortunately, the trend of the market follows their predictions. The purpose of example 3 is to demonstrate that any traders with different strategies and future anticipations can find an appropriate way of employing financial *option* insurance. In other words, trading strategies can be significantly enriched with the introduction of financial *option* insurance into the real practice.

# 3. Business Strategy of Matching

An appropriate business strategy plays a crucial role for the *third entity* to minimize its potential losses, and at the same time, to maximize its profitability. As mentioned above, one crucial component of financial *option* insurance is to determine the appropriate yardstick. It is suggested that the yardstick (in the previous examples, a yardstick of 50% is assumed), which is employed to determine the *insurance premium* and the reimbursement amount, should be tweaked according to the prevailing financial market conditions. Therefore, the *third entity* can immediately respond to the rapidly changing situations.

Under the perfect market assumption, there is always a balance between two equal-sized parties of *calls* and *puts* such that the plenty number of financial *derivative* investors can be protected by financial *option* insurance, by either entering into a new insurance contract or purchasing an existing insurance contract from the *secondary market of insurance contracts* hosted by the *third entity*. By following such an assumption, no matter whether the price of the *underlying security* (stock price) ascends or plunges at maturity, under the current version of reimbursement clauses, it always allows the *third entity* to stand in a profiting position. It is because the *third entity* only reimburses one of these paired investors, whose unexercised *option* is *OTM*, and stands a chance to profit from the entire *insurance premium* if their *options* are *ATM* at maturity. Such a strategy is depicted in Figure 8(a).

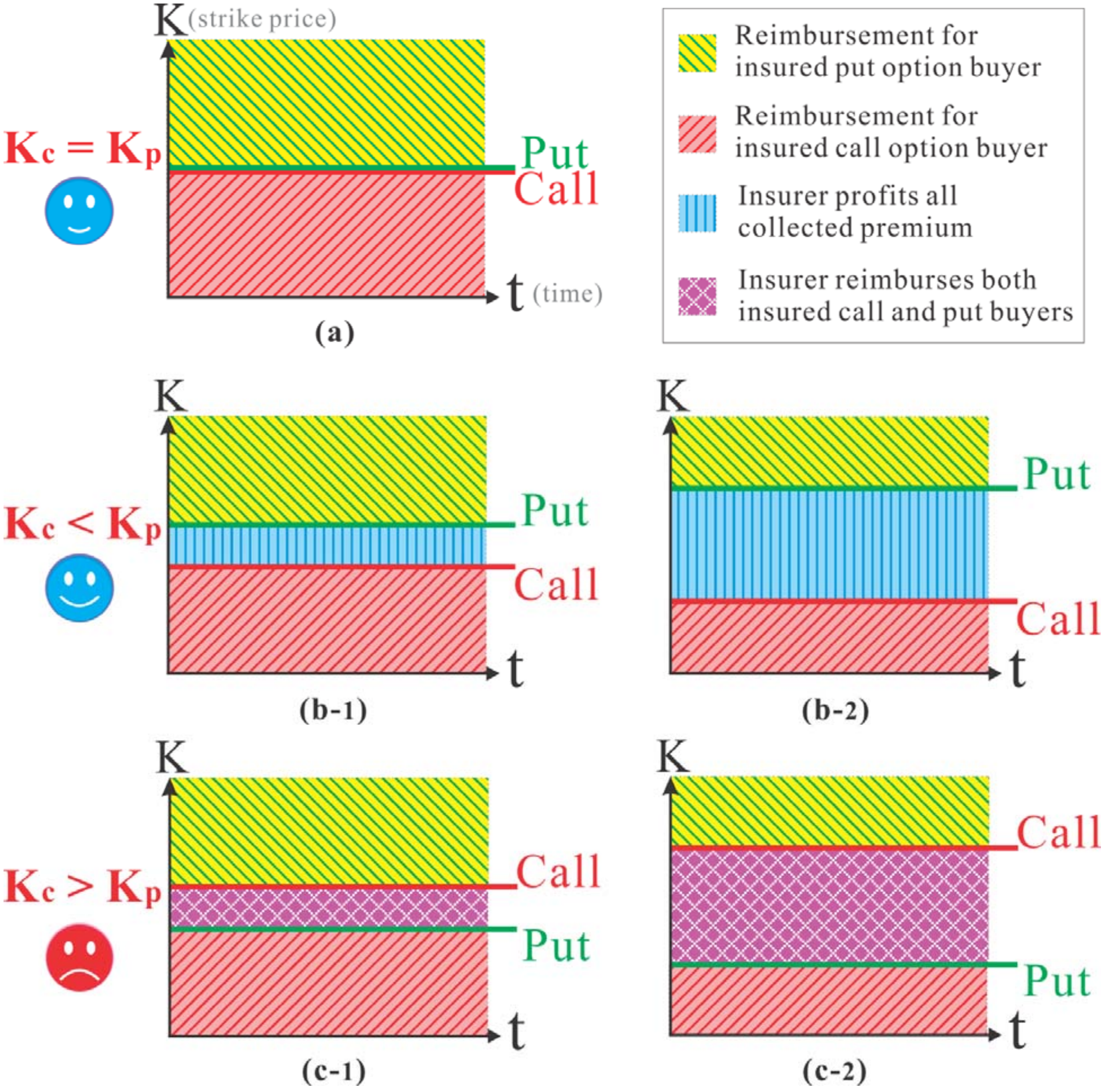

**Figure 8:** *Three possible scenarios of matching between different moneyness of calls and puts*

However, such balance rarely exists as there are always the unequal-sized parties of *calls* and *puts* with different moneyness, due to various expectations about future and dissimilar adopted strategies. As a straight consequence, only a small portion of investors, who seek the protections on their *option premium*, can be benefited, leaving the majority of the investors pending on the *matching system*.

The situation can be improved with the addition of the secondary market, which enables *option* investors to purchase an existing insurance contract to protect their positions. The intrinsic characteristic of the financial *option* insurance (*i.e.*, the new insurance contract is issued based upon the existing long position of the corresponding *option*) limits the supply of new insurance contracts into the secondary market. As a consequence, the inadequate supply of insurance contract increases the general price level of insurance contracts. Hence, *option* investors may abandon the protection for their position due to the excessive *insurance premium* in contrast to the corresponding reimbursement amount.

In order to enable the investors with *calls* and *puts* of different moneyness to be protected by the issued *option* insurance, a specifically designed mathematical model is proposed in this paper. It has two portions: the business strategy of matching and a verification-and-modification process. The proposed model enables the *option* investors with *calls* and *puts* of different moneyness to be allocated as paired investors. Meanwhile, it minimizes the exposure of *option insurer*'s position to any potential losses. In addition to that, the model enables the sufficient supply of insurance contracts with different moneyness to the *secondary market of insurance contracts*.

## I. Three possible scenarios of matching

The matching between the *calls* and *puts* of different moneyness, is likely to fall into one of the following three scenarios: 1) the *strike price* of *call option* equals to that of *put option*, *i.e.*, $K_C = K_P$; 2) the *strike price* of *call option* is lower than that of *put option*, *i.e.*, $K_C < K_P$; and 3) the *strike price* of *call option* is higher than that of *put option*, *i.e.*, $K_C > K_P$.

The first scenario, where $K_C = K_P$, is the ideal case for the *third entity* as it is under the perfect market assumption. As shown in Figure 8(a), there are two overlapping horizontal lines – the green line (above) represents the *strike price* of the *put option* and the red line (below) stands for the *strike price* of the *call option*. These two lines separate the entire quadrant into two regions – the yellow region (above the green line) and the pink region (below the red line). The concept is that if the separate price of the *underlying security* of financial *option* with various *maturity dates* (along the horizontal axis) remains in either yellow or pink region, the *third entity* reimburses insured *put* or *call option* buyer, respectively. However, if the price of *underlying security* at any *maturity date* of *option* is placed right on these two overlapping lines, the *third entity* is going to profit the entire *insurance premiums* collected from both insured *call* and *put option* investors, who have purchased the financial *option* insurance.

The second scenario, where $K_C < K_P$, is the most preferable scenario in the perspective of the *third entity*. As compared with the first scenario, there is one additional region – the blue region, which is covered by neither yellow nor pink region, as indicated in Figure 8(b-1). If the price of the *underlying security* at any specified *maturity date* falls into this region, the *third entity* profits all *insurance premium* collected from both *call* and *put option* investors. Under the same scenario, if the difference is increased between these two *strike prices*, *call* and *put*, as shown in Figure 8(b-2), the blue region is enlarged accordingly. It implies that the *third entity* stands a much higher chance to profit all collected *insurance premium* without reimbursing any single party of paired investors.

The third scenario, where $K_C > K_P$, is a completely reversed version of the second scenario. The yellow region overlaps the pink region in the creation of a purple region between these two price lines, as shown in Figure 8(c-1). On the contrary, if the price of the *underlying security* at any *maturity date* along the horizontal axis falls into the purple region, it may simply spell a catastrophic consequence for the *third entity* as it has to reimburse both insured *call* and *put option* investors. The situation becomes severe if the moneyness difference between these two *options*, *call* and *put*, is increased, as shown in Figure 8(c-2).

The three possible scenarios of matching between two *options* of different moneyness suggest that the most preferable scenario to *option insurer* should be the second one. The level of preference is proportional to the difference between two *strike prices* of the corresponding *options*. On the other hand, the least desirable scenario to the *third entity* is the third one, which becomes severe if the difference between two *strike prices* is increased. The business strategy of matching is therefore converted to the determination of optimal pairs of *call* and *put option* investors. The objective is to minimize the occurrence of the third scenario and maximize the occurrence of the second scenario.

## II. Mathematical model

Based upon three different scenarios of matching between *calls* and *puts* with different moneyness, it is possible to draft a simple business strategy of matching among the *calls* and *puts* of different moneyness. It begins by separating a group of investors, who demand for financial *option* insurance to protect their naked positions, into two subgroups of totally reversed reimbursement clauses. For each *call option* of certain *strike price*, the *third entity* ranks its preference of matching to the *put options* with different *strike prices* in accordance with three possible scenarios of matching. Similarly, for each *put option* of certain *strike price*, the *third entity* also ranks its preference of matching to the *call options* of different *strike prices* with respect to three possible scenarios.

Therefore, the initial case is streamlined into a problem of determining a stable assignment between *calls* and *puts* with assigned rankings. Furthermore, the rankings of preference can be tabulated in terms of a matrix, called *ranking matrix*. Each entry of ranking matrix consists of two numbers, let us say, $i, j$. The first number, $i$, is the ranking of *put options* given by the *third entity* in correspondence to each *call option*. The second number, $j$, is the ranking of *call options* given by the *third entity* in correspondence to each *put option*.

Once the ranking matrix is established, the remaining procedures of matching become relatively simple. First, all *call options* match to the *put options* in terms of their first preference, *i.e.*, $i = 1$, in the ranking matrix. Based on the second number of each entry, *put options* retain the highest ranking which matches to it, and reject the remaining. Second, all unassigned *call options* match the *put options* in terms of their second preference, *i.e.*, $i = 2$. Such an acceptance-and-rejection procedure repeats until all *call options* have been allocated to a *put option*. After the completion of the assignment process, it returns with a primary matching list.

The *option insurer* (or the *third entity*) does not propose the *option* insurance contracts based upon such a matching list directly. Although the matching list indicates the most optimal combinations of *calls* and *puts* with different moneyness, it does not guarantee that the outcome is satisfying from the perspective of the *option insurer*, that is, the frequencies of the second scenario is always higher than that of the third scenario. And even if the frequencies of the second scenario is larger than or equivalent to that of the third scenario, it does not assure that the *third entity* stands a much higher chance to gain all *insurance premiums* collected from both *call* and *put option* investors (*i.e.*, the price of *underlying security* is placed on the blue region on *maturity date* under the second scenario, as shown in Figure 8(b) than to reimburse both parties (*i.e.*, the price of underlying asset is placed on the purple region on *maturity date* under the third scenario, as shown in Figure 8(c).

Therefore, before finalizing the insurance contract, the matching list is subjected to a simple verification-and-modification process, which aims to ensure the total amount, collected from both *call* and *put option* investors under the second scenario, exceeds the total amount that is used to reimburse both parties under the third scenario. For paired investors under the second scenario in pair $m$, the corresponding *strike price* on maturity for these paired *call* and *put* investor are $K_{C,m}$ and $K_{P,m}$, respectively, and $K_{P,m} > K_{C,m}$, where $m$ is a non-negative integer. Similarly, for the paired investors under the third scenario in pair $n$, the *strike price* on maturity for these paired *call* and *put* investor are $K_{C,n}$ and $K_{P,n}$, respectively, and $K_{P,n} < K_{C,n}$, where $n$ is a non-negative integer. The difference between $K_{C,m}$ and $K_{P,m}$ reflects the chance for *option insurer* profiting the entire pre-collected *insurance premium* ($R_m$) in pair $m$. A larger gap stands for a higher chance and vice versa. On the other hand, the difference between $K_{C,n}$ and $K_{P,n}$ reveals the probability for *option insurer* suffering a loss ($L_n$) by paying both paired investors in pair $n$. In order to comparing the relative occurrence of these two events, a normalized factor $D$ is introduced, where $D = \max\{K_{P,m} - K_{C,m}, K_{C,n} - K_{P,n}\}$. The resultant weighting equation is

$$W = \frac{1}{D}\left[\sum_m \left(K_{P,m} - K_{C,m}\right)R_m + \sum_n \left(K_{P,n} - K_{C,n}\right)L_n\right]. \qquad (\#)$$

If the weightage value $W > 0$, the *option insurer* is likely to accept the matching list and to propose the insurance contract to the corresponding *option* investors. Otherwise, the pair $n$ under the third scenario with the maximal absolute value of $K_{P,n} - K_{C,n}$ is rejected and the result is undergoing another round of verification-and-modification process until $W > 0$. These rejected pairs are pending on the waiting list of the *matching system* for another round of allocation.

## III. Example

Assuming that there are a group of eight *option* investors, half of them are *call option* investors {A, B, C, D}, and another half are *put option* investors, {α, β, γ, δ}. To protect their naked *option* against potential losses, all of them decide to purchase financial *option* insurance from the *third entity*. The *options*, *calls* and *puts*, purchased by the corresponding investors, are of different moneyness as shown in Figure 9(a).

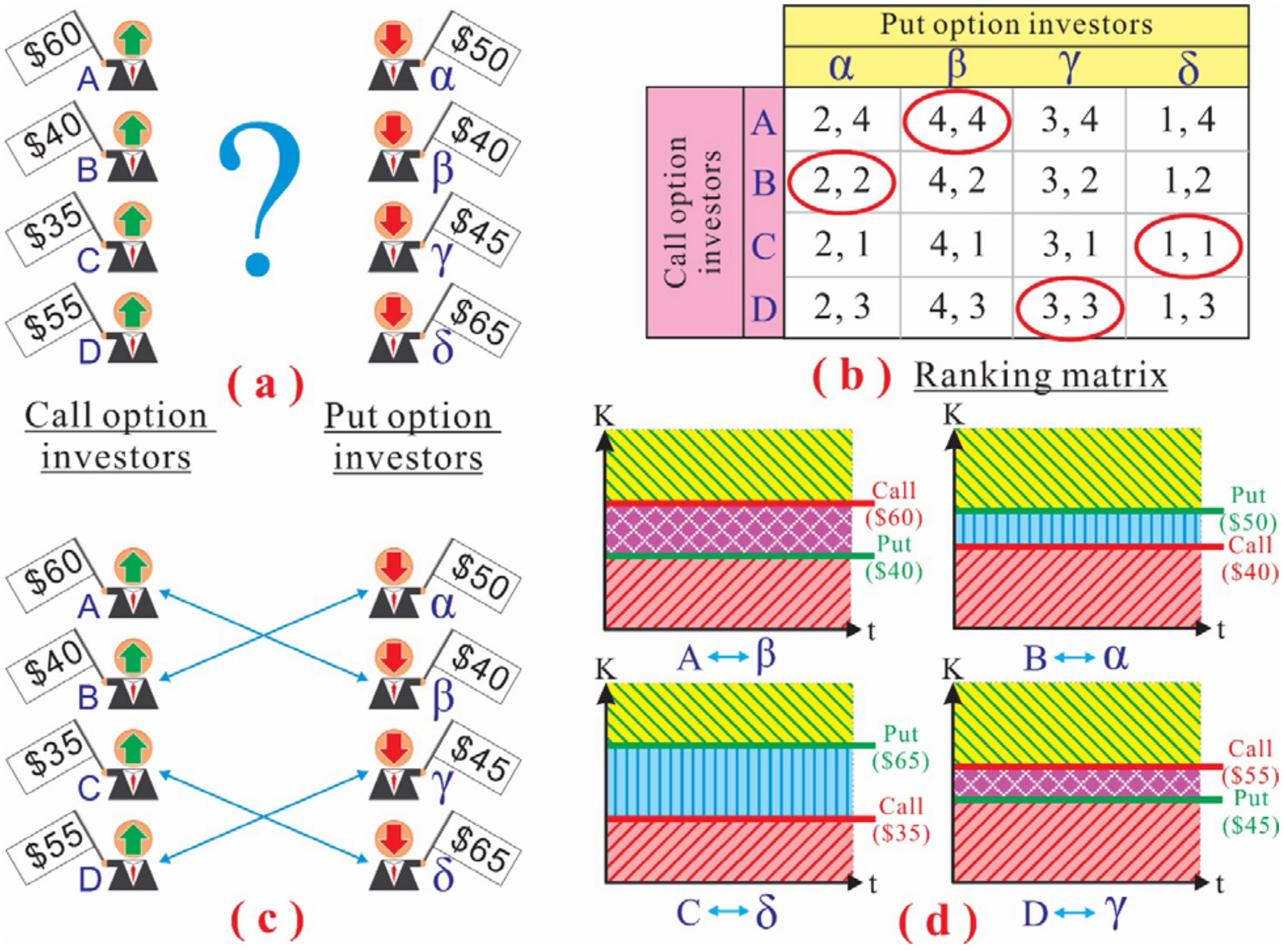

**Figure 9:** *Demonstration of matching strategy among option investors*

The very initial task for the *third entity* is to identify the most optimal matching among these investors before proposing an appropriate insurance contract. It begins by ranking the preference of matching in the perspective of each *call option* investor to the *put option* investors based on the result obtained from three possible scenarios of matching. Relying on three possible scenarios of matching, it is possible to convert the problem into a ranking matrix as shown in Figure 9(b).

The ranking assignment procedure is straightforward; let us take *put option* investor α as an example. *Option* investor α holds a *put* of $50 (*strike price*). When he combines to the *call option* investors A and D with a *call* of $60 and $55, the situation is obviously unfavorable to the *third entity* as the *strike price* of *call* is higher than that of *put* (the third scenario). In these two matches, {A α} pair is much worse as the price difference is larger than that of {D α}. Comparably, the combinations of {B α} and {C α} are better than the previous two matches as the *strike price* of *put* is higher than that of *call* (the second scenario). In addition, {C α} is better than {B α} as the price difference is higher than that of {B α}. Therefore, from the perspective of the *third entity*, the final ranking for *put option* investor α is likely to be {A, B, C, D} = {4, 2, 1, 3}, which is indicated in the first column of the ranking matrix, Figure 9(b). After the completion of ranking matrix, by repeating acceptance-and-rejection procedure as mentioned above, it is capable of determining the most optimal matching in the combination of {A β, B α, C δ, D γ}, as shown in Figure 9(c). The developed business strategy of matching is analogous to the generalized Tian Ji's horse racing strategy [18–20] and the Nobel prize-winning stable allocation theory [21,22]. In our case, the preferences can easily be quantified in terms of moneyness of financial *option*.

The primary matching list {A β, B α, C δ, D γ} undergoes a verification-and-modification process. In order to demonstrate the concept, it is possible to estimate the conceptual *option premium* of these *options* by entering the following information into an *option* price calculator: spot price of *underlying security*: $50, days until expiration: 180 days, annualized interest rate: 1%, dividend yield: 0, volatility: 40%. The obtained *option premium* prices are displayed in Table.

| **Set** | **Calls** | | | **Puts** | | |
|---|---|---|---|---|---|---|
| | Call investor | Strike price | Option premium | Put investor | Strike price | Option premium |
| *m* | | | | | | |
| 1 | B | $40 | $11.7 | α | $50 | $5.4 |
| 2 | C | $35 | $15.7 | δ | $65 | $16.2 |
| *n* | | | | | | |
| 1 | A | $60 | $2.4 | β | $40 | $1.5 |
| 2 | D | $55 | $3.8 | γ | $45 | $3.1 |

**Table:** *Corresponding strike price and option premium of paired option investors*

Based upon the yardstick of 50% and the service charge of 1% for the reimbursement and all *option* investors holding the minimal amount of the corresponding *option*, the value of $R_m$ and $L_n$ can be determined as follows: $R_1$ = 5.85, $R_2$ = 8.1, $L_1$ = 1.188, $L_2$ = 1.881. Thereafter, it is possible to substitute all related values into equation (#). The resultant weightage value, $W$, can be obtained as follows:

$$W = \frac{1}{30}\left[(50-40)5.85 + (65-35)8.1 + (40-60)1.188 + (45-55)1.881\right] = 8.631 > 0$$

As the weightage value $W > 0$, the matching list is acceptable in the perspective of the *option insurer*, that is, the *third entity* stands a much higher chance to gain all *insurance premiums* collected from both *call* and *put option* investors than to reimburse both parties. By combining the business strategy of matching as well as the verification-and-modification process, the desired outcome can be achieved for the *option insurer* when matching among the investors with the *calls* and *puts* of different moneyness.

# 4. Option Insurance in Financial Markets

The feasibility of financial *option* insurance can be addressed in terms of three aspects: market acceptance, *risk* profile and profitability of *option insurer*, and positive market effect.

## I. Market acceptance

Financial *option* insurance integrates two distinct concepts of financial *options* and insurance. In prevailing financial market, there is a two-entity structure composing *option writers* and *holders*. *Option writer* sells a right to the *option holder* in exchange for the *option premium*, expecting that the *option* expires worthlessly on maturity. Although, in contrast to *option writers*, whose *risk* is theoretically unlimited as the price of *underlying security* may soar by unlimited amount (for *call option writers*) or plunge to zero (for *put option writers*), the *risk* encountered by *option holders* is limited to the *option premium* paid. It is always optimal to have their *option premium* being compensated at a cost of sacrificing a relatively small portion of potential profits as investors have either little or no control over the profitability from a particular set of *portfolio*. This is just analogous to an individual who purchases accident insurances and has either little or no control over the probability of accident occurring to him.

In contemporary financial market, there are several existing strategies, which attempt to limit the *risk* (and return) participations. The simplest versions are *protective* and *synthetic puts*. The *protective put* is a bullish *portfolio* strategy whereas the investor buys the shares of one particular stock, and at the same time, purchases the *put option* to cover their shares. Such a strategy enables an investor to enjoy potentially unlimited returns, as the stock price may theoretically soar to infinity, while limiting the downside stock price movement, due to the nature of the *put option*. The strategy of *protective put* is demonstrated in example 2 (Section 2III, the strategy of investor **C**). The *synthetic put*, on the other hand, is a bearish *portfolio* strategy where the investor short sells one stock, and at the same duration, purchases *call option* to cover the position of short-selling. The resultant *portfolio* enables investor to enjoy potentially significant returns, as the stock price may theoretically plunge to zero, while limiting the upside stock price movement, due to the nature of the *call option*. The strategy of *synthetic put* is depicted in example 2 (Section 2III, the strategy of investor **D**).

By comparing the potential profit/loss graph for *protective* and *synthetic puts* (as shown in Figure 6) and that of *call* and *put purchase* (as shown in Figure 2), it is possible to figure out the strategies of *protective* and *synthetic puts* are very alike to the strategies of *call* and *put purchase*, respectively, as all of these strategies successfully limit the *risk* on one-sided stock price movement while exposing the position to potentially significant returns if the anticipation is correct. The *protective put*, in contrast to *call purchase*, offers superior protections to adverse market movements as the investor can hold the purchased stock until the most favorable selling point. As a trade-off, the strategy demands significant capital involvement (in longing the corresponding stock), and hence, notably limits the potential returns if the prediction is right. Similarly, for *synthetic put* strategy, although unlike the *protective put* strategy, which demands an investor to pay the full amount, it still requires the investor to maintain the minimum marginal requirements. Therefore, in the strategy of *synthetic put*, the investor can hold in the short position of stock, to limit the potential *risk* at the cost of fulfilling the marginal requirements, which also limits its potential returns, as compared to the strategy of *put purchase*.

In addition to the *option* strategies of *protective* and *synthetic puts*, there are many other available *option* strategies, which are frequently adopted by investors to limit their potential *risk*, such as *collars*, *debit bull call spread*, *debit bear put spread*. These strategies generally involve one shorting position on *option*, which provides immediate amount to compensate the cost of opening other long positions. They successfully limit the capital involvement (in contrast to the strategies of *protective* and *synthetic puts*); on the other hand, the potential returns are also greatly restricted. Let us take the strategy of *debit bull call spread* as an example. *Debit bull call spread* is a bullish strategy, which combines a long position on the *call option* with a lower *strike price* and a short position on the identical *call option* with a higher *strike price*. The maximal profit is limited to a fixed amount, that is, the difference between the *higher strike price* and the *lower strike price* less the *net debit of spread*, which is minuscule as compared to the potential returns of *protective* and *synthetic puts*.

Is there a conceivable method, which exposes investors to theoretically unlimited returns at a reasonable price tag? One feasible solution is introducing one additional entity, the *third entity*, into the establishment of a three-entity structure. It is economic sense for market participants to seek out an avenue of reducing the *risk* (*option premium* invested in the financial *options*) at a limited cost (*insurance premium*). The primal idea of the business is tantamount to spread the total *risk* between paired investors by mitigating the cost of investors' *option* with a pre-paid *insurance premium*. If *an option premium* is affordable in view of possible refund, they are more than willing to purchase financial *option* insurance to insure against their potential losses due to "misfortunate bet".

In financial market, the annual trading volume of financial *derivatives* amounts to trillions of dollars, and therefore, it is certainly a huge market to tap. In addition, there is no noteworthy evidence to suggest that the financial *option* insurance with any rudimentary intention of an "insurance policy" to *derivative* market participants has been developed so far.

At first glance, financial *option* insurance and *rebate barrier option*, a type of *exotic option*, have some similarities. It is because the rebate can be deemed, from the perspective of *option* investor, as a compensation for the losses if the *knock-in option* remains inactive or the *knock-out option* ceases to exist within the life time of the *option*. In reality, the working principles of these two financial products are completely different.

The major difference is lying on the corresponding trigging mechanism. The *rebate barrier option* is path dependent. For the *knock-out barrier option* with rebate, once the barrier level is breached by the instantaneous price movement of *underlying security*, the *option* is terminated, and the rebate is payable at the time of the event or on the *maturity date* to the *option holder* depending on specified terms. Even if the price of underlying asset moving in a completely reversed direction, which is favorable to the *option* investor, the *option* does not come into existence after breaching the barrier level; hence, such a *barrier option* prohibits the investor from exposing its position to potential profit thereafter. Similarly, for the *knock-in barrier option* with rebate, once the barrier level is breached, the *option* comes into exist. Even if the subsequent trend of the *underlying security* moves in an unfavorable direction in the perspective of the *option* investor, its position may no longer be protected. In summary, the trigging mechanism of *rebate barrier option* is rigid in the perspective of the investor as it does not allow the corresponding adjustment of its strategic position in response to the price movements of the *underlying security* during the life time of the *option*.

Conversely, financial *option* insurance provides higher flexibility to the *option* investor as it is path independent. The trigging mechanism is fixed, which is occurred only on the *expiration date* of the insurance contract (*i.e.*, the *maturity date* of the underlying *option*). It only checks whether the underlying *option* remains unexercised and *OTM*. If it does, the insured *option* investor is allowed to claim for reimbursement from the *option insurer*. Otherwise, the insurance contract expires worthlessly. Moreover, the financial *option* and the insurance contract are not bounded together (*i.e.*, the insurance contract remains valid even if the underlying *option* is exercised or traded for the profit within the life time of the insurance contract). In addition, the ownership of an insurance contract can be traded separately on the platform regulated by the *third entity*. In such arrangement, it provides the investor with a higher degree of freedom to adjust its position in response to the price movements of the underlying asset. Such flexibility cannot be delivered by the *rebate barrier option*.

From the macroscopic point of view, the *rebate barrier option* and financial *option* insurance are operated in a different manner. The transaction of *rebate barrier option* consists of two parties: the *option writer* and investor. From the perspective of *option holder*, the *rebate barrier option* generally demands a lower premium as compared to the identical *vanilla option* in exchange for a less favorable condition when it is executed (*i.e.*, the *option* may cease to exist or does not spring into exist within the life time of the *option*). Although a rebate, usually a fraction of the *option premium*, is payable to the investor when an unfavorable condition occurs, such rigid trigging mechanism, as mentioned above, greatly limits the strategic choices of the investor. Moreover, it is because the *rebate barrier option* is not a standardized contract (*i.e.*, it cannot be traded on an exchange), seeking a potential buyer on the secondary market of such an *option* is much more challenging as compared to the corresponding *vanilla option*. On the other hand, the barrier level can be deemed as an effective *risk* management technique for the *option writer*. However, in a real trading scenario, the terms of rebating are very rarely employed by the *option writer* as it diminishes its potential profit.

Conversely, the transaction of financial *option* insurance involves three parties: *call option* investor, *put option* investor, and *option insurer* (also known as the *third entity*). For *option* investors, the insurance contracts as well as the underlying *options* are standardized contracts, which can be actively traded on the corresponding platforms. The primary role of *option insurer* (or the *third entity*) is selling *option* insurance in pairs, that is, these paired *option* investors have completely reversed expectations. Unlike the *risk* management method adopted by the *writer* of *rebate barrier option*, who is directly responsible for compensating the losses suffered by the *option* investor, a fraction of the profit earned by the *option* investor with the right "bet" is utilized to mitigate the loss of investor with the wrong "bet". The role of *option insurer* is alike to a financial intermediary dedicated to matching and redistributing the *risk* among *option* investors without having any direct *risk* participations.

## II. Risk profile and profitability

Any *risk*, which lasts sufficiently long, may be assessed and insured. The phenomenon explains why the *option insurer*'s strategy is employed to financial *derivatives* instead of shares, as its value is highly unpredictable. In essence, financial *option insurer* is completely different from any existing financial intermediates participating in the contemporary financial *option* market. It is because its ability for reimbursement is to shift *risk* from one group of investors to another group of investors by redistributing the losses among participating members.

In prevailing financial market, there is a two-entity structure composing *option writers* and *holders*. Brokerage firms are the major financial intermediates, who exercise trading inquires on behalf of market participants, and profit from obtaining the commissions of various transactions. Its associated *risk* is mainly coming from the *option writers*, whose position is vulnerable to theoretically unlimited *risk*. The *option writer* may fail to deliver the underlying asset or to accept the delivery of the underlying asset as stipulated in the insurance contract, which is referred to as counterparty *risk*. It arises because of the credit *risk* of the *option writer* [23]. The *risk*, on most occasions, can be minimized by financially strong intermediaries, which utilize collateral or netting arrangement to make good on the trade. However, as outlined in [24], such arrangement is unable to completely eliminate the *risk*. And the *risk* can become much severer when in a major financial crash, as the huge number of defaults may overwhelm even the strongest brokerage firms.

The financial *option insurer*, or the *third entity*, as a newly proposed financial intermediate, follows a completely distinct working principle. Therefore, it is associated with a different *risk* profile and profitability.

The *option insurer* is very alike to an insurance company dedicated to financial *option* market. The issuing of new insurance contract requires investors to open a long position on either *call option* or *put option*. In addition to that, the new insurance contract is only issued in pairs – a *call option* buyer and a *put option* buyer with the same expiry date and moneyness. Such matching mechanism ensures that at most one party of these paired investors (or ultimate insurance contract *holder* if the insurance contract is obtained from the secondary market) is legal to claim reimbursement from the *option insurer*. The unqualified party has to let the insurance contract expire worthlessly. Besides, upon the successful issuing of a new insurance contract, the determined *insurance premium* is collected from both paired investors. Meanwhile, the reimbursement clauses specifies the compensation condition (ultimate insurance contract *holder* has to identify holding the specified unexercised *OTM option* on expiry date) and the maximal compensation amount (in Section 2IV, compensation amount is \$11.88 per share for *call option* and \$7.425 per share for *put option*). Therefore, in the perspective of the *third entity*, such mechanism confines the maximal potential loss of *option insurer* to the entire pre-collected *insurance premium* (when the price of *IKEA* fall below \$500), while standing a chance to retain partial *insurance premium* (when the price of *IKEA* rises above \$500) or even the complete amount (when the price of *IKEA* remains at \$500) as its earnings. In summary, potential return for *option insurer* is always greater than zero, which is determined by the difference between the collected *insurance premium* and reimbursing amount, and perfectly independent of the corresponding market trend.

This nature is not affected by the involvement of the *secondary market of insurance contracts* as the same mechanism perfectly insulates the *option insurer* from any potential *risk* induced by the price movements of insurance contracts in the secondary market. In addition, permitting insurance contracts to be actively traded in the secondary market hosted by *option insurer* have four major implications. 1) It enables any *option* market participants to seek protections by purchasing the insurance contract from the secondary market, in the case of that the *matching system* cannot immediate allocate paired *option* investors or the offered reimbursement clauses is not attractive as compared to the existing insurance contracts listed in the secondary market. Moreover, as mentioned in Section 2I, unlike issuing a new insurance contract, purchasing an existing insurance contract does not require an investor to open a long position on the underlying *option*. Therefore, it allows investors to secure an insurance contract prior entering a long position on the corresponding *option*. 2) It facilitates the insured investors, who leave the position prior to expiry, to have their *insurance premium* being compensated or even subjected to a gain. 3) The information on the transaction price of insurance contracts in the secondary market reflects various expectations of the investors. It can be employed by *option insurer* as a reference in adjusting the yardstick determined by the pricing structure to stand in a profiting position. As such information is transparent to all entities involved in the *option* market, one additional factor, which skews the behavior of the *option writer*, should be taken into account. For instance, if the spot price of one particular financial *option* insurance in the secondary market is close to its reimbursement amount, it simply reflects that the market participants anticipate such an *option* is likely to be *OTM* on maturity. Such expectation may further diminish the *option premium* obtained by *option writers* (*i.e.*, *option writers* have to lower the *option premium* to maintain its attractiveness to the potential *option* buyers); hence, inhabiting the willingness on selling such an *option* as the *risk option premium* becomes too low to compensate for the potential *risk* associated with the position. Conversely, if the spot price of one particular insurance contract in the secondary market is far away from its reimbursement amount as specified in the insurance contract, such an *option* has a higher chance to remain *ITM* at maturity. Therefore, *option writers* may demand a higher *option premium* to compensate its potential *risk*. Under this scenario, it may stimulate *option writers* to open the position due to excessive *risk option premium*. 4) The transaction fee can be collected upon each successful transfer of ownership of insurance contract (assuming that the transaction fee charged by the *third entity* is 2% of the market value of the insurance contract, the earnings from the transaction fee alone is $250 in total, as indicated in Section 2IV).

In a real financial market, however, the perfect market rarely exists as there are always the unequal-sized "camps" of *calls* and *puts* with different *strike prices*. This is mainly caused by different expectations about the performance of the underlying asset and adopted trading strategies. When encountering the imperfect reality, the proposed business strategy of matching makes it possible in the allocation of different *calls* and *puts* with various moneyness, thus effectively restraining the potential *risk* encountered by the financial *option insurer*, maximizing its exposure to the potential profits, enabling *option* investors to seek immediate protections, and supplying sufficient liquidity to the *secondary market of insurance contracts*.

Moreover, *option insurer* can also act as a normal insurance company to invest its collected *option premium* into fixed-income financial products, or even take the certain amount of *risk* in providing insurance for the unequal-sized "camps" during the issuing of a new insurance contract.

### III. Positive market effect

As covered by financial *option* insurance, investors are stimulated to participate in the *option* market. From a macroscopic perspective, *option insurer* reduces the overall *risk* of financial *derivative* market and promotes benign market involvement. Reimbursement stems mainly from the *option premium* prepaid to the *option insurer* upon the issuing of insurance contract. Therefore, cash flow between *option insurer* and insured *option* investor is completely independent of prevailing financial *derivative* market. Moreover, the matching mechanism, a verification-and-modification process, insurance policies (reimbursement condition and amount), the detachment of issuing and the *secondary market of insurance contracts*, radically reduce the credit *risk* of *option insurer*; hence, there is no solvency issue for the *third entity* to repay the stipulated amount to the compliant insurance contract *holder*, whose unexercised *option* is *OTM* on maturity. Therefore, even in the event of masse early exercise of *options* during a crisis, the robustness of *option insurer* remains unchallenged.

The established business model of the *third entity* allows the practice of pure speculating on insurance contracts in the secondary market (as specified in Section 2IV, the corresponding strategy adopted by *speculator* **G**), which may increase the overall instability to the market. However, such motivation is somewhat suppressed by the insurance policies, that is, the *speculators* have to provide the evidence of holding the specified unexercised *OTM options* on reimbursement date. Even if the *speculators* may obtain the required *OTM options* from the exchange by paying very low *option premium* (due to the vanished time value of the *option*). The potential profit is still greatly limited by stipulated reimbursement amount (for instance, as specified in example 3, the reimbursement amount for the *call* and *put options* is \$12 and \$7.5 per share, respectively, without counting the service charge for reimbursement). In addition to that, the *speculators* may choose to sell the insurance contract on the secondary market prior to expiry. Comparing to the strategy of the *option purchase*, the time value of insurance contract, on most occasions, is lower than that of the corresponding financial *option*. It is because the *option* can be exercised without fulfilling any requirements prior to the *maturity date*. Therefore, introducing the *secondary market of insurance contracts* does not provide much incentive for speculating on the insurance contract. The addition of the secondary market, on the other hand, does advocate the practice of hedging the potential *risk* associated with the existing position. For instance, investor **H** combines an *ITM put option* with a *put option* insurance obtained from the secondary market, which allows exposing his position to limited *risk* (*i.e.*, the price of *underlying security* failing in the range of \$496 − \$500) and standing on a profiting position whereas the price goes sideways.

The business of *option insurer* is seemingly an arbitrager but its profitability is much lower than a lucrative bet by an *option* investor. Therefore, the position may, in a certain degree, foreclose the deterioration of financial *option* insurance business.

This proposed business idea seeks to supplement the prevalent market practices of *option* and insurance business where the contemporary financial market does not have avenues for market participants to seek insurances in protecting their *option premiums* against “misfortune” or “wrong bet”, especially for the *option speculators* with an entirely unprotected position.

The central idea of financial *option* insurance is unique and distinctive in which it brings two starkly dissimilar concepts together, and merges them in the creation of a completely new situation where opportunities arise for a *third entity* to exist. Moreover, it creates an overall win-win situation for all entities involved by redistributing *risk* among a larger pool of investors. With the potential of lower *risk* and having a payoff that outweighs its investment, an investor is positively enticing to introduce the financial *option* insurance into their existing positions.

# 5. Concluding Remarks

A completely new conceptual model of financial *option* insurance is proposed. It integrates two starkly dissimilar concepts of financial *derivative* (*option*) and insurance by introducing one additional entity, known as the *third entity*, into the contemporary two-entity financial structure. In order to enable the investors with *calls* and *puts* of different moneyness to be protected by the issued *option* insurance, a specifically designed mathematical model is proposed in this paper. It has two portions: the business strategy of matching and a verification-and-modification process.

The *third entity* seeks to replicate insurance practices into the field of *risk* management in finance, whereas *risk* is distributed among a larger pool of entities of *speculators* or *portfolio managers*. This would not be made possible without a complete understanding of the process flow of how *speculators* and *portfolio managers* utilize *options* to speculate or to hedge against adverse market movements and how insurance industries generally operate. This leads to the successful integration of concepts of insurance policies and financial *derivatives*.

It is a worthwhile strategy if the *third entity* is sufficiently large to have simultaneously many investors taking opposite position to each other in terms of the *option purchase* and the level of *option* moneyness. Even under imperfect market conditions, the proposed business strategy of matching shall allow more entities to be protected by the financial *option* insurance while minimizing the exposure of the *third entity* to the potential losses.

In addition, the *third entity* hosts the issuing market and the secondary market of financial *option* insurance, which has a tremendous impact on contemporary financial *option* market from macroscopic view. First, it reduces the overall *risk* of the prevailing financial *derivative* market *via* spreading the total *risk* among various entities. Second, it stimulates the willingness of market participation with the introduction of additional protection to their positions and enriches their trading strategies with the involvement of financial *option* insurance. Last but not least, the disclosure of *option* buyers, who seek for financial *option* insurance, always reflects a better sense of market expectations, which can even be treated as an effective index to monitor the overall performance of special underlying asset.

# Acknowledgements

This work was supported by the Ministry of Education of Singapore (M4011179) and Nanyang Technological University (M4081942).